# Emirati-Accented Speaker Identification in each of Neutral and Shouted Talking Environments

Ismail Shahin[1,a], Ali Bou Nassif[2,b], Mohammed Bahutair[3,c]

[1, 2, 3]Department of Electrical and Computer Engineering, University of Sharjah, Sharjah, UAE

[a]ismail@sharjah.ac.ae, [b]anassif@sharjah.ac.ae, [c]mohammed.bahutair@hotmail.com

**Abstract**

This work is devoted to capturing Emirati-accented speech database (Arabic United Arab Emirates database) in each of neutral and shouted talking environments in order to study and enhance text-independent Emirati-accented "speaker identification performance in shouted environment" based on each of "First-Order Circular Suprasegmental Hidden Markov Models (CSPHMM1s), Second-Order Circular Suprasegmental Hidden Markov Models (CSPHMM2s), and Third-Order Circular Suprasegmental Hidden Markov Models (CSPHMM3s)" as classifiers. In this research, our database was collected from fifty Emirati native speakers (twenty five per gender) uttering eight common Emirati sentences in each of neutral and shouted talking environments. The extracted features of our collected database are called "Mel-Frequency Cepstral Coefficients (MFCCs)". Our results show that average Emirati-accented speaker identification performance in neutral environment is 94.0%, 95.2%, and 95.9% based on CSPHMM1s, CSPHMM2s, and CSPHMM3s, respectively. On the other hand, the average performance in shouted environment is 51.3%, 55.5%, and 59.3% based, respectively, on "CSPHMM1s, CSPHMM2s, and CSPHMM3s". The achieved "average speaker identification performance in shouted environment based on CSPHMM3s" is very similar to that obtained in "subjective assessment by human listeners".

**Keywords**: Emirati-accented speech database; hidden Markov models; "neutral talking environment"; "shouted talking environment"; speaker identification; suprasegmental hidden Markov models.## 1. Introduction and Literature Review

"Speaker recognition" is categorized into two different main types: "speaker identification and speaker verification (authentication)". "Speaker identification" is defined as the method of automatically finalizing who is speaking from a group of known speakers. "Speaker verification" is defined as the method of automatically accepting or rejecting the identity of the claimed speaker. "Speaker identification" can be heavily utilized in investigating criminals to conclude the speculated suspects who produced a voice captured at the episode of a crime. On the other side, "speaker verification" is broadly utilized in security entry to services through a telephone such as: "home shopping, home banking transactions using a telephone network, security control for restricted information areas, remote access to computers, and many other telecommunication services" [1]. "Speaker recognition" is grouped, based on the text to be spoken, into "text-dependent and text-independent cases". In the "text-dependent case", "speaker recognition" necessitates the speaker to utter speech for the same text in both training and testing phases, while in the "text-independent case", "speaker recognition" is independent on the text being uttered.

Arabs can communicate among themselves in the Arab countries in one of the four regional dialects of the Arabic language. These dialects are: Egyptian (*e.g.* Egyptian), Levantine (*e.g.* Palestinian), North African (*e.g.* Algerian), and Gulf Arabic (*e.g.* Emirati) [2].

In the areas of speaker recognition and speech processing and recognition, most of the research work has been focused on speech spoken in English language [1], [3], [4], [5] while very limited number of studies focus on these areas on speech uttered in Arabic language [6], [7], [8], [9], [10]. One of the reasons of these few number of studies is the small number of accessible Arabic speech datasets in these areas [11], [12]. Al-Dahri *et.al* [6] studied word-dependent "speaker identification systems" encompassing 100 speakers speaking Arabic isolated words based on "Hidden Markov Models (HMMs)". "Mel-Frequency Cepstral Coefficients (MFCCs)" have been adopted as the extracted features of the utilized dataset. They reported 96.3% accuracy to recognize the correct speaker [6]. Krobba *et.al* [7] investigated the effect of GSMEFR speech data on the performance of a "text-independent speaker identification system" based on "Gaussian Mixture Models (GMMs)" as classifiers. The recognition assessment was also performed using original "ARADIGIT sampled at 16 KHz and its 8 KHz down-sampled version". The "ARADIGIT database" is made up of 60 speakers generating the ten Arabic digits with a replicate of three times each. Various experiments were accomplished to calculate the deterioration caused by diverse aspects of the simulated codec [7]. Mahmood *et.al* [8] proposed and implemented novel features called "Multi-Directional Local Feature (MDLF)" for speaker recognition. In order to extract MDLF, a windowed speech signal has been processed based on "Fast Fourier Transform (FFT)" and passed through 24 Mel-scaled Filter Bank, followed by a log compression stage. MDLF carries the characteristics of the speaker in time spectrum and yields an improved performance. As a classifier, GMM with a different number of mixtures has been used in their work. Their results showed that the proposed MDLF gave higher recognition accuracy than the traditional MFCC features. The MDLF obtains outstanding results both in "text-dependent and text-independent speaker recognition" using Arabic and English databases [8]. Saeed and Nammous [9] studied a speech-and-speaker (SAS) identification system based on spoken Arabic digit recognition. The speech signals of the Arabic digits from zero to ten have been processed graphically (the signal has been considered as an object image for additional processing). The identification and classification stages have been conducted with "Burg's estimation model" and the algorithm of "Toeplitz matrix minimal eigenvalues" have been used as the major tools for signal-image description and feature extraction. In the classification stage, both "conventional and neural-network-based" methods have been used. Their reported average overall success rate was 97.5% to recognize one uttered word and recognizing its speaker, and 92.5% to recognize a three-digit password (three individual words) [9]. Tolba [10] used "Continuous Hidden Markov Models (CHMMs)" as a classifier to automatically identify Arabic speakers from their voices. MFCCs have been utilized as the extracted features of speech signals. Ten Arabic speakers have been used to assess his proposed CHMM-based engine. His reported speaker identification performance is 100% and 80% for "text-dependent and text-independent" systems, respectively [10].

There are few number of publications that use Emirati-accented speech database [13], [14]. To the best of our knowledge, there are only two available studies in the areas of speech and speaker recognition that use Emirati-accented database [13], [14]. Shahin and Ba-Hutair [13] studied in one of their work Emirati-accented "speaker identification systems" in a neutral talking environment based on each of "Vector Quantization (VQ), GMMs, and HMMs" as classifiers. The Emirati database is made up of 25 men and 25 women Emirati native speakers. These speakers uttered 8 famous Emirati sentences that are broadly utilized in the "United Arab Emirates society". The eight sentences were neutrally uttered (no stress or emotion) by each speaker 9 times with a span of 1 – 3 seconds. They used MFCCs as the extracted features of their dataset. Their results showed that "VQ" is superior to each of "GMMs and HMMs" for both "text-dependent and text-independent" Emirati speaker identification. In another work by Shahin [14], he focused on evaluating a "text-independent speaker verification" using Emirati speech dataset collected in a neutral talking environment. The dataset was captured from 25 men and 25 women Emirati native speakers who uttered 8 commonly-used Emirati sentences. MFCCs have been utilized as the extracted features of speech signals. Three distinct classifiers have been employed in his work. These classifiers are: "First-Order Hidden Markov Models (HMM1s), Second-Order Hidden Markov Models (HMM2s), and Third-Order Hidden Markov Models (HMM3s)". His results showed that HMM3s outperform each of HMM1s and HMM2s for a text-independent Emirati-accented speaker verification.

This work aims at collecting Emirati-accented speech database (Arabic United Arab Emirates database) uttered in each of "neutral and shouted talking environments" in order to study and enhance text-independent Emirati-accented "speaker identification performance" in a shouted environment based on each of "First-Order Circular Suprasegmental Hidden Markov Models (CSPHMM1s), Second-Order Circular Suprasegmental Hidden Markov Models (CSPHMM2s), and Third-Order Circular Suprasegmental Hidden Markov Models (CSPHMM3s)" as classifiers. These classifiers are novel for Emirati-accented speaker identification. In addition, seven experiments have been conducted to thoroughly study Emirati-accented "speaker identification in each of neutral and shouted talking environments". In this research, our speech database was captured from fifty Emirati native speakers (twenty five male and twenty five female) uttering eight common Emirati sentences in each of "neutral and shouted environments". MFCCs have been adopted as the extracted features of our collected dataset. This work is different from our two previous studies [13], [14]. In [13], Shahin and Ba-Hutair studied Emirati-accented "speaker identification systems in a neutral environment" only based on each of "VQ, GMMs, and HMMs" as classifiers. In [14], Shahin focused on evaluating a text-independent speaker verification using Emirati-accented speech database captured in a neutral environment only based on three different classifiers: "HMM1s, HMM2s, and HMM3s".

The remaining structure of this paper is as follows: Brief overview of suprasegmental hidden Markov models is given in Section 2. The basics of "CSPHMM1s, CSPHMM2s, and CSPHMM3s" are given in Section 3. Section 4 gives the details of the captured speech database used in this work and the extracted features. "Speaker identification algorithm in each of neutral and shouted environments based on CSPHMM1s, CSPHMM2s, and CSPHMM3s" and the experiments are discussed in Section 5. Section 6 gives the results attained in the present work and their discussion. Finally, concluding remarks are given in Section 7.

## 2. Overview of Suprasegmental Hidden Markov Models

In many of his studies, Shahin exploited and tested SPHMMs as classifiers. Such studies are: "speaker identification in each of emotional and shouted environments" [4], [15], [16], speaker verification in emotional environments [17], and emotion recognition [18], [19]. SPHMMs have proven to be superior models over HMMs in these studies since SPHMMs possess the capability to summarize some states of HMMs into a new state named "suprasegmental state". "Suprasegmental state" is able to look at the observation sequence through a larger window. Such a state lets observations at rates that fit the case of modeling emotional and stressful signals. Prosodic information cannot be recognized at a rate that is utilized for acoustic modeling. The prosodic features of a unit of emotional and stressful signals are named "suprasegmental features" because they affect all the segments of the unit signal. Prosodic events at the levels of "phone, syllable, word, and utterance" are characterized using "suprasegmental states", while "acoustic events" are patterned using "conventional hidden Markov states".

"Prosodic information and acoustic information" are merged together within HMMs as [20],

$$\text{“}\log P\left(\lambda^v, \Psi^v | O\right) = (1-\alpha). \log P\left(\lambda^v | O\right) + \alpha. \log P\left(\Psi^v | O\right) \quad (1)$$

where $\alpha$ is a weighting factor. When:

$$\begin{cases} 0.5 > \alpha > 0 & \text{biased towards acoustic model} \\ 1 > \alpha > 0.5 & \text{biased towards prosodic model} \\ \alpha = 0 & \text{biased completely towards acoustic model and no effect of prosodic model} \\ \alpha = 0.5 & \text{not biased towards any model} \\ \alpha = 1 & \text{biased completely towards prosodic model and no impact of acoustic model} \end{cases} \quad (2)$$

$\lambda^v$ is the $v^{th}$ acoustic model, $\Psi^v$ is the $v^{th}$ SPHMM model, $O$ is the observation vector of an utterance, $P\left(\lambda^v | O\right)$ is the probability of the $v^{th}$ HMM model given the observation vector $O$, and $P\left(\Psi^v | O\right)$ is the probability of the $v^{th}$ SPHMM model given the observation vector $O$".

Further information about SPHMMs can be obtained from the references [21], [22].

# 3. Basics of CSPHMM1s, CSPHMM2s, and CSPHMM3s

## 3.1. "First-order circular suprasegmental hidden Markov models"

"CSPHMM1s" have been derived from "acoustic First-Order Circular Hidden Markov Models (CHMM1s)". Zheng and Yuan proposed and applied CHMM1s for "speaker identification in neutral environment" [23]. Shahin demonstrated that CHMM1s lead "Left-to-Right First-Order Hidden Markov Models (LTRHMM1s) for speaker identification in shouted environment" [24]. More information about CHMM1s can be found in references [23], [24].

Fig. 1 illustrates an example of an essential topology of CSPHMM1s that has been formed from CHMM1s. As an example, this figure consists of six "first-order acoustic hidden Markov states: $q_1, q_2, ..., q_6$" placed in a circular shape. $p_1$ is a "first-order suprasegmental state" that is made up of "$q_1, q_2$, and $q_3$". $p_2$ is a "first-order suprasegmental state" that is comprised of "$q_4, q_5$, and $q_6$". $p_1$ and $p_2$ are "two suprasegmental states" placed in a circular style. $p_3$ is a "first-order suprasegmental state" which consists of $p_1$ and $p_2$.

## 3.2. "Second-order circular suprasegmental hidden Markov models"

"CSPHMM2s" have been obtained from acoustic "Second-Order Circular Hidden Markov Models (CHMM2s)" [22]. Shahin proposed, used, and assessed CHMM2s for speaker recognition in each of "shouted and emotional environments" [24]. Shahin showed in his studies that these models outperform each of "LTRHMM1s, LTRHMM2s, and CHMM1s" since CHMM2s hold the characteristics of both "CHMMs and HMM2s" [24]. Readers can get additional details about CHMM2s and CSPHMM2s from reference [24] and [22], respectively.

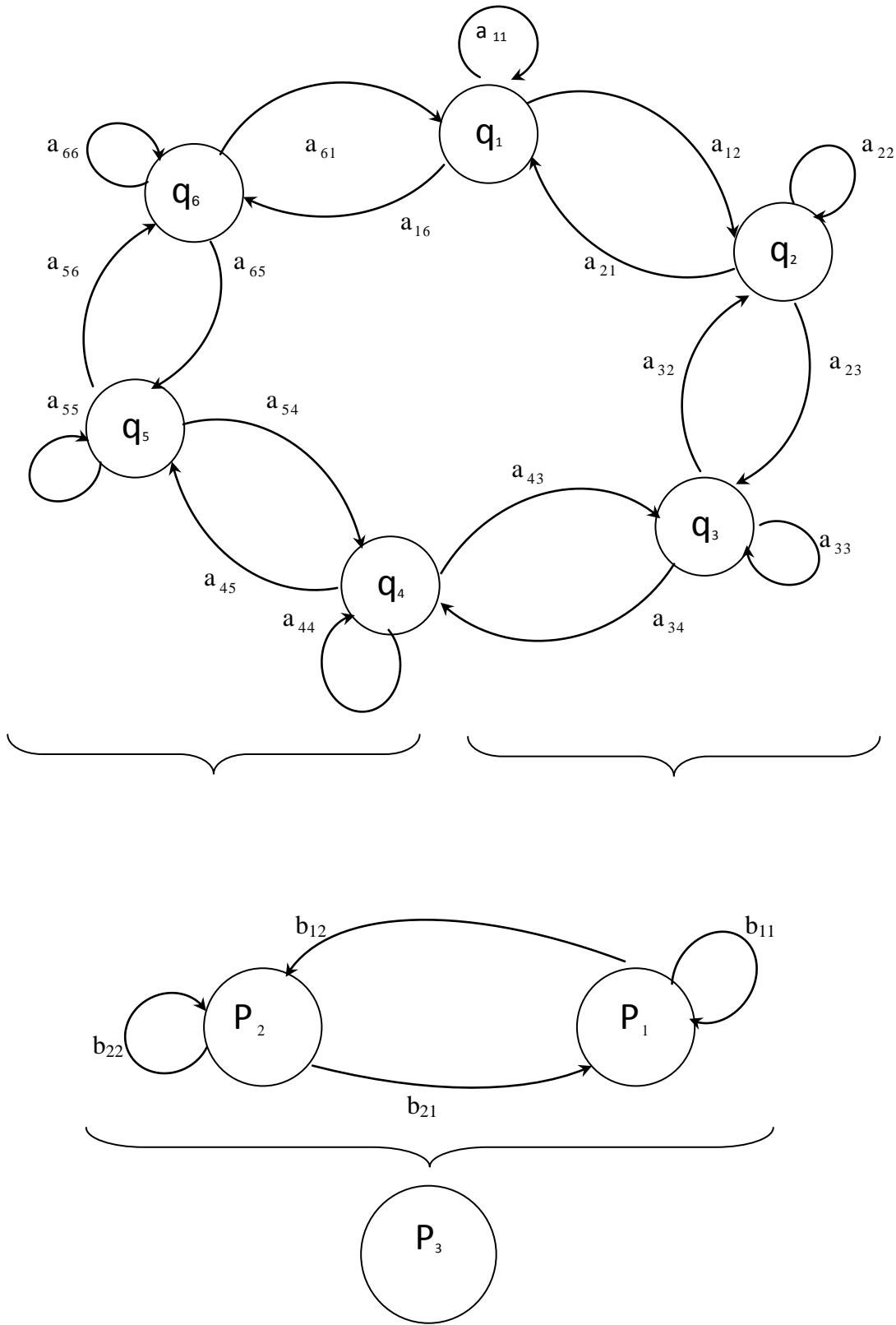

**Figure 1.** Basic structure of CSPHMM1s attained from CHMM1s

As an example of CSPHMM2s, the six "first-order acoustic circular hidden Markov states" of Fig. 1 are replaced by six "second-order acoustic circular hidden Markov states" ordered in the similar shape. $p_1$ and $p_2$ become "second-order suprasegmental states" located in a circular form. $p_3$ is a "second-order suprasegmental state" which is comprised of $p_1$ and $p_2$.

### 3.3. "Third-order circular suprasegmental hidden Markov models"

"Third-Order Circular Suprasegmental Hidden Markov Models" have been structured from acoustic "Third-Order Hidden Markov Models (HMM3s)". In one of his studies, Shahin [25] employed, utilized, and tested HMM3s to improve low "text-independent speaker identification performance in shouted environment". He showed that HMM3s are superior to each of "HMM1s and HMM2s" in such an environment [25].

In HMM1s, the "underlying state sequence is a first-order Markov chain" where the stochastic process is stated by a 2-D matrix of a "priori transition probabilities" ($a_{ij}$) between states $s_i$ and $s_j$ where $a_{ij}$ is given as [25],

$$a_{ij} = \text{Prob}\left(q_t = s_j \big| q_{t-1} = s_i\right) \qquad (3)$$

In HMM2s, the "underlying state sequence is a second-order Markov chain" where the stochastic process is expressed by a 3-D matrix ($a_{ijk}$). Hence, the "transition probabilities in HMM2s" are given as [25],

$$a_{ijk} = \text{Prob}\left(q_t = s_k \big| q_{t-1} = s_j, q_{t-2} = s_i\right) \qquad (4)$$

In HMM3s, the "underlying state sequence is a third-order Markov chain" where the stochastic process is stated by a 4-D matrix ($a_{ijkw}$). Accordingly, the "transition probabilities in HMM3s" are given as [25],

$$a_{ijkw} = \text{Prob}\left(q_t = s_w \mid q_{t-1} = s_k, q_{t-2} = s_j, q_{t-3} = s_i\right) \quad (5)$$

The "probability of the state sequence", $Q \triangleq q_1, q_2, ..., q_T$, is expressed as:

$$\text{Prob}(Q) = \Psi_{q_1} a_{q_1 q_2 q_3} \prod_{t=4}^{T} a_{q_{t-3} q_{t-2} q_{t-1} q_t} \quad (6)$$

where "$\Psi_i$ is the probability of a state $s_i$ at time $t = 1$, $a_{ijk}$ is the probability of the transition from a state $s_i$ to a state $s_k$ at time $t = 3$". $a_{ijk}$ can be computed from equation (4). Thus, the initial parameters of HMM3s can be obtained from the trained HMM2s.

"Prosodic and acoustic information" within CHMM3s can be mingled into CSPHMM3s as [16],

$$\text{"}\log P\left(\lambda^v_{CHMM3s}, \Psi^v_{CSPHMM3s} \mid O\right) = (1-\alpha) \cdot \log P\left(\lambda^v_{CHMM3s} \mid O\right) + \alpha \cdot \log P\left(\Psi^v_{CSPHMM3s} \mid O\right) \quad (7)$$

where, $\lambda^v_{CHMM3s}$ is the acoustic third-order circular hidden Markov model of the $v^{th}$ speaker and $\Psi^v_{CSPHMM3s}$ is the suprasegmental third-order circular hidden Markov model of the $v^{th}$ speaker".

To give an example of CSPHMM3s, the six "first-order acoustic circular hidden Markov states" of Fig. 1 are substituted by six "third-order acoustic circular hidden Markov states" structured in the identical figure. $p_1$ and $p_2$ come to be third-order suprasegmental states positioned in a circular structure. $p_3$ is a third-order suprasegmental state which is comprised of $p_1$ and $p_2$.

Shahin [16] showed that CSPHMM3s outperform each of CSPHMM1s and CSPHMM2s for speaker identification in a shouted environment. This is because the characteristics of CSPHMM3s are composed of the characteristics of both "Circular Suprasegmental Hidden Markov Models (CSPHMMs)" and "Third-Order Suprasegmental Hidden Markov Models (SPHMM3s)".

## 4. Emirati-Accented Speech Dataset and Extraction of Features

### 4.1. Collected Emirati-Accented Speech Dataset

In this study, twenty five male and twenty five female native Emirati speakers with ages spanning from 14 to 55 year old uttered the Emirati-accented speech database (Arabic database). Each speaker uttered 8 common Emirati sentences that are heavily utilized in the "United Arab Emirates society". The eight sentences were portrayed by every speaker in every "neutral and shouted environments" (two isolated environments) 9 times with a span of 2 – 5 seconds. These speakers were untrained to utter the Emirati sentences in advance to avoid any overstated expressions (to make this database spontaneous). The total collected number of utterances was "5400 ((50 speakers × first 4 sentences × 9 repetitions/sentence in neutral environment for training session) + (50 speakers × last 4 sentences × 9 repetitions/sentence × 2 talking environments for testing sessions))". The sentences are tabulated in Table 1 (the right column gives the sentences in Emirati accent, the left column demonstrates the English version, and the middle column shows the phonetic transcriptions of these sentences). This dataset was captured in two separate and distinct sessions: training session and testing (identification) session.

The recorded dataset was captured in an uncontaminated environment in the College of Communication, "University of Sharjah, United Arab Emirates" by a group of professional engineering students. The database was collected by a "speech acquisition board using a 16-bit linear coding A/D converter and sampled at a sampling rate of 44.6 kHz". The signals were then down sampled to 12 kHz. The samples of signals were pre-emphasized and then segmented into slices (frames) of 20 ms each with 31.25% intersection between successive slices.

Table 1

Emirati speech database in its: English version, phonetic transcriptions, and Emirati accent

| No. | English Version | Phonetic Transcriptions | Emirati Accent |
|---|---|---|---|
| 1. | We will meet with you in an hour. | / bintlɑ:ga wıjɑ:k ʕugub sɑ:ʕah / | بنتلاقى وياك عقب ساعة |
| 2. | Go to my father he wants you. | /si:r ʕınd abu:jeh yibɑ:k / | سير عند ابويه يباك |
| 3. | Bring my cell phone from the room. | /hɑ:t tilıfu:ni: mınıl ḥıjrah / | هات تيلفوني من الحجرة |
| 4. | I am busy now I will talk to you later. | / maʃɣɔ:ɫ(a) ʌḥi:n baramsık ʕʌb sɑ:ʕəh / | مشغول/مشغولة الحين برمسك عقب |
| 5. | Every seller praises his market. | / kıl byaiʕ yımdeḥ su:gah / | كل بياع يمدح سوقه |
| 6. | A stranger is a wolf whose bite wounds won't heal. | / ılġari:b ði:b w ʕaẓitah maṭi:b / | الغريب ذيب و عضته ما تطيب |
| 7. | Show respect around some people and show self-respect around other people. | / nɑ:æs ıḥʃımhom w nɑ:s ıḥʃım nafsak ʕanhom / | ناس احشمهم و ناس احشم نفسك عنهم |
| 8. | Don't criticize what you can't get and don't swirl around something you can't obtain. | / illi magdart tiyibah lɑ: tʕi:bah w illi mɑ:ṭu:lah lɑ: ṭḥu:m ḥu:lah / | اللي ما قدرت تييبه لا تعيبه و اللي ما تطوله لا تحوم حوله |

### 4.2. Extraction of Features

As features extracted from the Emirati-accented speech database, MFCCs have been utilized in this study as the appropriate features that extract the phonetic content of Emirati-accented signals. These features have been largely used in many areas of speech. Examples of such areas are speech and speaker recognition. MFCCs have proven to outperform other coefficients in the two areas and they have shown to grant a high-level approximation of human auditory perception [17], [26], [27], [28]. In this study, a 32-dimension feature analysis of MFCCs (16 static MFCCs and 16 delta MFCCs) was utilized to found the observation vectors in each of "CSPHMM1s, CSPHMM2s, and CSPHMM3s". In each one of these models, a "continuous mixture observation density" was chosen. In every suprasegmental model, the number of "conventional states, $N$, is nine and the number of suprasegmental states is three (every suprasegmental state is made up of three conventional states)".

### 5. "Speaker Identification Algorithm in each of Neutral and Shouted Environments Based on CSPHMM1s, CSPHMM2s, and CSPHMM3s" and the Experiments

The "training phase" of each of "CSPHMM1s, CSPHMM2s, and CSPHMM3s" is very similar to the "training phase" of the "conventional CHMM1s, CHMM2s, and CHMM3s", respectively. In the "training phase" of each of "CSPHMM1s, CSPHMM2s, and CSPHMM3s" (completely three isolated training phases), "suprasegmental first-order circular models, suprasegmental second-order circular models, and suprasegmental third-order circular models" are trained on top of "acoustic first-order circular models, acoustic second-order circular models, and acoustic third-order circular models", respectively. In every "training phase", the $v^{th}$ speaker model has been obtained utilizing the "first four sentences" of the Emirati-accented speech dataset with 9 replicates for every sentence spoken by the $v^{th}$ speaker in "neutral environment". The entire number of utterances that have been utilized to obtain the $v^{th}$ speaker model in every "training phase" is 36 ("first 4 sentences × 9 repetitions/sentence").

In the "test (identification) phase" of each of "CSPHMM1s, CSPHMM2s, and CSPHMM3s" (entirely three separate "test phases"), every one of the "fifty speakers" independently utters each sentence of the "last four sentences" of the dataset (text-independent) with 9 replicates per sentence in each of "neutral and shouted environments". The overall number of utterances that have been uttered in every identification phase per talking environment is 1800 "(50 speakers × last 4 sentences × 9 repetitions/sentence)". The probability of producing every utterance per speaker is independently calculated based on each of "CSPHMM1s, CSPHMM2s, and CSPHMM3s". For every one of these three "suprasegmental models", the model with the maximum probability is selected as the output of "speaker identification" as given in the coming formula for every talking environment:

$$V^* = \arg\max_{50 \geq v \geq 1} \left\{ P\left(O \mid \lambda^v_{model}, \Psi^v_{model}\right) \right\} \tag{8}$$

where "$O$ is the observation vector or sequence that corresponds to the unknown speaker, $\lambda^v_{model}$ is the acoustic hidden Markov model (this model can be one of: CHMM1s, CHMM2s, or CHMM3s) of the $v^{th}$ speaker and $\Psi^v_{model}$ is the suprasegmental hidden Markov model (this model can be one of: CSPHMM1s, CSPHMM2s, or CSPHMM3s) of the $v^{th}$ speaker".

# 6. Results and Discussion

The current research focuses on collecting Emirati-accented speech database uttered in each of neutral and shouted environments for the purpose of studying and improving text-independent Emirati-accented "speaker identification in shouted environment" based on each of "CSPHMM1s, CSPHMM2s, and CSPHMM3s" as classifiers. These classifiers are novel for Emirati-accented speaker identification. In this research, the weighting factor ($\alpha$) has been selected to be equal to 0.5 to avoid biasing towards either "acoustic or prosodic model" in each of "CSPHMM1s, CSPHMM2s, and CSPHMM3s".

In this work, our Emirati-accented speech database was collected in "one of the studios that belong to the College of Communication at the University of Sharjah in the United Arab Emirates". Twenty five speakers per gender volunteered to utter eight common Emirati sentences with a replication of nine times each under each of neutral and shouted environments. There were some difficulties that faced the team who captured this database:

1) The collected database is acted and unspontaneous. Therefore, the achieved results based on using such data are biased. The vast majority of studies usually use acted speech database since it is very difficult to collect spontaneous one.

2) There are some logistic problems in bringing old people to the university to collect their voices.

3) The speakers are volunteers and unprofessional ones.

Table 2 represents "speaker identification performance in each of neutral and shouted environments" using the Emirati-accented speech corpus based on each of the suprasegmental models: "CSPHMM1s, CSPHMM2s, and CSPHMM3s" as classifiers. This table clearly demonstrates that "speaker identification performance" is almost ideal in neutral environment based on each one of these three classifiers. The reason is that each acoustic model ("CHMM1s, CHMM2s, and CHMM3s") results in a high "speaker identification performance" in such an environment as given in Table 3. However, the performance has been steeply deteriorated in shouted environment since each corresponding acoustic model results in a poor "speaker identification performance" under this environment as shown in Table 3. It is evident from Table 2 that "CSPHMM3s" outperform each of "CSPHMM1s and CSPHMM2s" in shouted environment by 15.6% and 6.8%, respectively.

Table 2

"Speaker identification performance in each of neutral and shouted environments" using Emirati-accented dataset based on each of the suprasegmental models: "CSPHMM1s, CSPHMM2s, and CSPHMM3s"

| Suprasegmental model | Gender | "Speaker identification performance (%)" | |
|---|---|---|---|
| | | "Neutral environment" | "Shouted environment" |
| CSPHMM1s | Male | 95.2 | 52.1 |
| | Female | 92.8 | 50.5 |
| | Average | 94.0 | 51.3 |
| CSPHMM2s | Male | 96.0 | 56.7 |
| | Female | 94.4 | 54.3 |
| | Average | 95.2 | 55.5 |
| CSPHMM3s | Male | 96.6 | 60.4 |
| | Female | 95.2 | 58.2 |
| | Average | 95.9 | 59.3 |

Table 3

"Speaker identification performance in each of neutral and shouted environments" using Emirati-accented dataset based on each of the acoustic models: "CHMM1s, CHMM2s, and CHMM3s"

| Acoustic model | Gender | "Speaker identification performance (%)" | |
| --- | --- | --- | --- |
| | | "Neutral environment" | "Shouted environment" |
| CHMM1s | Male | 91.4 | 38.0 |
| | Female | 90.2 | 35.4 |
| | Average | 90.8 | 36.7 |
| CHMM2s | Male | 91.9 | 45.1 |
| | Female | 90.7 | 42.7 |
| | Average | 91.3 | 43.9 |
| CHMM3s | Male | 92.8 | 50.2 |
| | Female | 91.6 | 48.8 |
| | Average | 92.2 | 49.5 |

"Speaker identification performance based on the three acoustic models (CHMM1s, CHMM2s, and CHMM3s)" has been significantly decreased in "shouted environment" compared to that in "neutral environment" as shown in Table 3. Acoustically, when speakers shout, air pressure inside the speaker's vocal tract is increased significantly. This increase creates a big cavity that enlarges vortices inside the vocal tract. Enlarging the vortices results in an increase in the generation of sound that intersects with the original sound [29]. Hence, the original speaker's sound is contaminated with other sounds. Consequently, the mismatch that exists between the "training session in neutral environment" and the "testing session in shouted environment" is increased. This increase in the mismatch negatively affects "speaker identification performance in shouted environment based on each one of the acoustic models: CHMM1s, CHMM2s, and CHMM3s".

To demonstrate whether "speaker identification performance" differences ("speaker identification performance" based on CSPHMM3s and that based on each of CSPHMM1s and CSPHMM2s in each of "neutral and shouted environments") are actual or just appear from statistical variations, a "statistical significance test" has been conducted. The "statistical significance test" has been performed based on the "Student's *t* Distribution test" as given by the following formula,

$$t_{model\ 1, model\ 2} = \frac{\overline{x}_{model\ 1} - \overline{x}_{model\ 2}}{SD_{pooled}} \quad (9)$$

where "$\overline{x}_{model\ 1}$ is the mean of the first sample (model 1) of size *n*, $\overline{x}_{model\ 2}$ is the mean of the second sample (model 2) of the same size, and $SD_{pooled}$ is the pooled standard deviation of the two samples (models)" given as,

$$SD_{pooled} = \sqrt{\frac{SD_{model\ 1}^2 + SD_{model\ 2}^2}{2}} \quad (10)$$

where "$SD_{model\ 1}$ is the standard deviation of the first sample (model 1) of size *n* and $SD_{model\ 2}$ is the standard deviation of the second sample (model 2) of the same size".

In this study, the "calculated *t* values" between "CSPHMM3s" and each of "CSPHMM1s and CSPHMM2s" in each of "neutral and shouted environments" using the Emirati-accented dataset are tabulated in Table 4. This table illustrates that each "calculated *t* value in neutral environment" is smaller than the "tabulated critical value $t_{0.05}$ = 1.645 at *0.05* significant level". In contrast, each "calculated *t* value in shouted environment" is greater than the "tabulated critical value $t_{0.05}$ = 1.645". Therefore, "CSPHMM3s" significantly outperform each of "CSPHMM1s and CSPHMM2s in shouted environment". It is apparent that "CSPHMM3s" are superior models to each of "CSPHMM1s and CSPHMM2s for speaker identification" since the "characteristics of CSPHMM3s" are comprised of the characteristics of both "CSPHMM1s and CSPHMM2s". In CSPHMM3s, the "state sequence is a third-order suprasegmental chain" where the stochastic process is stated by a "4-D matrix" since the "state-transition probability at time $t+1$ depends on the states of the suprasegmental chain at times: $t$, $t-1$, and $t-2$". Accordingly, the stochastic process that is stated by a "4-D matrix" yields greater "speaker identification performance" than that defined by either a "2-D matrix (CSPHMM1s) or a 3-D matrix (CSPHMM2s)". The dominance of "CSPHMM3s" over each of the other two models becomes insignificant in "neutral environment since the acoustic models: CHMM1s, CHMM2s, and CHMM3s" function well in such an environment as shown in Table 3.

Table 4

"Calculated $t$ values between CSPHMM3s and each of CSPHMM1s and CSPHMM2s in each of neutral and shouted environments" using Emirati-accented corpus

| | Calculated $t$ value | |
|---|---|---|
| "$t_{model\ 1,\ model\ 2}$" | "Neutral environment" | "Shouted environment" |
| "$t_{CSPHMM3s,\ CSPHMM1s}$" | 1.498 | 1.882 |
| "$t_{CSPHMM3s,\ CSPHMM2s}$" | 1.523 | 1.801 |

Table 2 states also that speaker identification performance for male Emirati speakers is greater than that for female Emirati speakers. Therefore, it can be concluded from this experiment that male Emirati speakers can be easily identified compared to female Emirati speakers. This conclusion is in agreement with the UAE culture where the speech of the UAE female speakers is very close (female speakers' speech cannot be easily recognized); however, the speech of the UAE male speakers is different (male speakers' speech can be easily recognized). Table 5 gives the "calculated *t* values" between male and female Emirati speakers based on each of "CSPHMM1s, CSPHMM2s, and CSPHMM3s in each of neutral and shouted environments" using the Emirati-accented database.

Table 5

"Calculated *t* values" between male and female Emirati speakers based on each of "CSPHMM1s, CSPHMM2s, and CSPHMM3s in each of neutral and shouted environments" using the Emirati-accented database

|  | "Calculated *t* value" | |
|---|---|---|
| t $_{male, female}$ | "Neutral environment" | "Shouted environment" |
| CSPHMM1s | 1.699 | 1.702 |
| CSPHMM2s | 1.717 | 1.793 |
| CSPHMM3s | 1.796 | 1.809 |

The "calculated *t* values" between each "suprasegmental model" and its belonging "acoustic model in each of neutral and shouted environments" using the Emirati-accented dataset are given in Table 6. This table clearly demonstrates that each "suprasegmental model" is superior to its corresponding "acoustic model" in each environment as each "calculated *t* value" in this table is higher than the "tabulated critical value $t_{0.05}$ = 1.645".

Table 6

"Calculated *t* values" between each "suprasegmental model" and its corresponding "acoustic model in each of neutral and shouted environments" using Emirati-accented dataset

|  | "Calculated *t* value" | |
|---|---|---|
| "t $_{sup.\ model,\ acoustic\ model}$" | "Neutral environment" | "Shouted environment" |
| "t $_{CSPHMM1s,\ CHMM1s}$" | 1.798 | 1.823 |
| "t $_{CSPHMM2s,\ CHMM2s}$" | 1.806 | 1.861 |
| "t $_{CSPHMM3s,\ CHMM3s}$" | 1.875 | 1.880 |

The obtained "speaker identification performance in a neutral environment" using Emirati-accented speech database (none of the previous studies have focused on "speaker identification in shouted environment") based on "CSPHMM3s" has been competed with that reported in some prior studies (non-Emirati Arabic database) in the same environment of:

1) Al-Dahri *et.al* [6] who studied word-dependent speaker identification systems containing 100 speakers uttering Arabic isolated words based on HMMs as classifiers and MFCCs as the adopted extracted features of their used database. They achieved 96.3% as a speaker identification performance [6]. Our achieved results based on CSPHMM3s are very close to their attained results.

2) Mahmood *et.al* [8] who proposed and applied novel features called MDLF for speaker identification. Based on MDLF as the extracted features and GMM as the classifier, they obtained 98.9% as a speaker identification performance. Our reported results based on CSPHMM3s are very alike to their achieved ones.

3) Tolba [10] who used CHMMs as classifiers to automatically identify Arabic speakers from their voices and MFCCs as the extracted features of speech signals. He obtained speaker identification performance of 80.0%. Our attained results in the current work outperform his results.

Seven extra experiments have been separately executed in this study to extensively evaluate the attained "speaker identification performance in each of neutral and shouted environments" using Emirati-accented database based on "CSPHMM3s". The seven experiments are:

1) Experiment 1: "Speaker identification performance" using the Emirati-accented speech dataset based on "CSPHMM3s" has been competed with that based on four distinct "state-of-the-art models and classifiers". The four classifiers and models are: "Gaussian Mixture Models (GMMs) [30], [31], Support Vector Machine (SVM) [32], [33], Genetic Algorithm (GA) [34], [35], and Vector Quantization (VQ)" [36], [37]. Table 7 demonstrates speaker identification performance using the Emirati-accented dataset based on each of "GMMs, SVM, GA, VQ, and CSPHMM3s". It is apparent from Table 2 and Table 7 that CSPHMM3s lead "GMMs, SVM, GA, and VQ" for Emirati-accented "speaker identification in a neutral environment" by 7.9%, 4.4%, 8.1%, and 6.6%, respectively. The two tables also show that "CSPHMM3s" are superior to "GMMs, SVM, GA, and VQ" by 62.9%, 28.6%, 70.4%, and 83.6%, respectively, for "speaker identification in a shouted environment".

Table 7
"Speaker identification performance in each of neutral and shouted environments" using Emirati-accented database based on each of "GMMs, SVM, GA, VQ, and CSPHMM3s"

| Models | Gender | "Speaker identification performance (%)" | |
|---|---|---|---|
| | | "Neutral environment" | "Shouted environment" |
| GMMs | Male | 91.1 | 54.6 |
| | Female | 86.7 | 18.2 |
| | Average | 88.9 | 36.4 |
| SVM | Male | 92.2 | 56.7 |
| | Female | 91.6 | 35.5 |
| | Average | 91.9 | 46.1 |
| GA | Male | 89.6 | 40.4 |
| | Female | 87.8 | 29.2 |
| | Average | 88.7 | 34.8 |
| VQ | Male | 91.1 | 39.3 |
| | Female | 88.9 | 25.3 |
| | Average | 90.0 | 32.3 |
| CSPHMM3s | Male | 96.6 | 60.4 |
| | Female | 95.2 | 58.2 |
| | Average | 95.9 | 59.3 |

2) Experiment 2: The three classifiers and models: CSPHMM1s, CSPHMM2s, and CSPHMM3s have been evaluated on the same eight sentences of our collected database but are uttered in this experiment by speakers talking in a formal Arabic language (not biased towards any dialect). This database is called a formal-accented speech database. In this experiment (text-independent), both the training and testing phases are comprised of the formal-accented database. The first four neutrally-uttered sentences of this formal-accented database are used in the training session, while the last four sentences of the formal-accented database are utilized in the identification session (one session for neutral environment and another separate session for shouted environment). Table 8 illustrates text-independent speaker identification performance based on each of CSPHMM1s, CSPHMM2s, and CSPHMM3s in each of neutral and shouted environments utilizing the formal-accented dataset. It is evident from Table 9 that "average speaker identification performance" based on training and testing our system using the Emirati-accented database is higher than that based on training and testing our system using the formal-accented database in each of neutral and shouted environments. This is because the eight sentences of our database fit better to native Arabic Emirati speakers rather than to nonnative Arabic Emirati speakers. These eight sentences are basically used in the daily communications among Emirati people only and they are not used among speakers talking in a formal Arabic language. Table 9 clearly shows that speaker identification improvement rate is more significant in shouted environment than in neutral environment.

Table 8

"Speaker identification performance in each of neutral and shouted environments" using formal-accented dataset based on each of "CSPHMM1s, CSPHMM2s, and CSPHMM3s"

| Models | Gender | "Speaker identification performance (%)" | |
|---|---|---|---|
| | | "Neutral environment" | "Shouted environment" |
| CSPHMM1s | Male | 91.1 | 47.4 |
| | Female | 90.3 | 45.6 |
| | Average | 90.7 | 46.5 |
| CSPHMM2s | Male | 92.8 | 50.2 |
| | Female | 91.2 | 48.6 |
| | Average | 92.0 | 49.4 |
| CSPHMM3s | Male | 92.9 | 55.0 |
| | Female | 92.1 | 53.0 |
| | Average | 92.5 | 54.0 |

Table 9

Speaker identification improvement rate of using Emirati-accented database over formal-accented database in each of "neutral and shouted environments based on each of CSPHMM1s, CSPHMM2s, and CSPHMM3s"

| Models | "Speaker identification improvement rate (%)" | |
|---|---|---|
| | "Neutral environment" | "Shouted environment" |
| CSPHMM1s | 3.6 | 10.3 |
| CSPHMM2s | 3.5 | 12.3 |
| CSPHMM3s | 3.7 | 9.8 |

3) Experiment 3: This experiment is the same as Experiment 2. However, the training phase is made up from the formal-accented speech database and the test phase is composed of the Emirati-accented speech database (one test session in neutral environment of the Emirati-accented database and another distinct test session in shouted environment of the same database). Therefore, in this experiment, we train our system with the formal-accented database and we test the system with the Emirati-accented database based on each of CSPHMM1s, CSPHMM2s, and CSPHMM3s. Speaker identification performance based on each of CSPHMM1s, CSPHMM2s, and CSPHMM3s in each of neutral and shouted environments trained using the formal-accented database and tested using the Emirati-accented database is given in Table 10. Table 11 clearly exemplifies that training and testing our system using the same Emirati-accented database is superior to that training with the formal-accented database and testing with the Emirati-accented database. This is because of the mismatch that exists between the training and testing phases. It is evident from this table that the superiority is significant in shouted environment while it is insignificant in neutral environment.

Table 10

"Speaker identification performance in each of neutral and shouted environments" using formal-accented database in the training phase and Emirati-accented database in the testing phase based on each of "CSPHMM1s, CSPHMM2s, and CSPHMM3s"

| Models | Gender | "Speaker identification performance (%)" | |
|---|---|---|---|
| | | "Neutral environment" | "Shouted environment" |
| CSPHMM1s | Male | 89.3 | 45.2 |
| | Female | 88.1 | 43.0 |
| | Average | 88.7 | 44.1 |
| CSPHMM2s | Male | 90.2 | 47.1 |
| | Female | 90.0 | 45.3 |
| | Average | 90.1 | 46.2 |
| CSPHMM3s | Male | 91.7 | 51.9 |
| | Female | 90.9 | 50.7 |
| | Average | 91.3 | 51.3 |

Table 11

Speaker identification improvement rate of using the same Emirati-accented database in both training and testing phases over that using formal-accented database in the training phase and Emirati-accented in the testing phase in neutral and shouted environments based on each of "CSPHMM1s, CSPHMM2s, and CSPHMM3s"

| Models | "Speaker identification improvement rate (%)" | |
|---|---|---|
| | "Neutral environment" | "Shouted environment" |
| CSPHMM1s | 6.0 | 16.3 |
| CSPHMM2s | 5.7 | 20.1 |
| CSPHMM3s | 5.0 | 15.6 |

4) Experiment 4: "Speaker identification performance in neutral environment" using the Emirati-accented database has been compared with one of our prior work [13]. In one of our previous studies, we focused on Emirati speaker identification systems in neutral environment based on each of VQ, GMMs, and HMMs as classifiers. These systems have been tested on our collected Emirati speech database which is comprised of 25 male and 25 female Emirati speakers using MFCCs as features. The database in [13] is different from the database being used in the current research. In [13], our database was captured from 25 men and 25 women Emirati speakers with ages between 14 and 27 years old. These speakers uttered 8 common Emirati sentences that are extensively utilized in the UAE society. Our results yield, for text-independent systems, average speaker identification performance of 94.58%, 86.6%, and 74.8%, based on VQ, GMMs, and HMMs, respectively. These results are smaller than those achieved based on CSPHMM1s (94.0%), CSPHMM2s (95.2%), and CSPHMM3s (95.9%).

5) Experiment 5: In this experiment, Emirati-accented speaker identification has been assessed for diverse values of $\alpha$ based on CSPHMM3s. Fig. 2 clarifies average speaker identification performance in each of neutral and shouted environments using the Emirati-accented database based on CSPHMM3s for various values of $\alpha$ (0.0, 0.1, 0.2,

…, 0.9, 1.0). It is apparent from this figure that as the value of $\alpha$ grows in the range [0,0.5], "speaker identification performance in shouted environment" increases significantly, while the performance insignificantly increases when α spans in the range [0.6,1.0]. This figure shows also that increasing the value of $\alpha$ has an insignificant effect on improving speaker identification performance in neutral environment. The conclusion that can be drawn in this experiment states that "suprasegmental hidden Markov models" have more influence than "acoustic models on speaker identification performance in shouted environment" of Emirati-accented database.

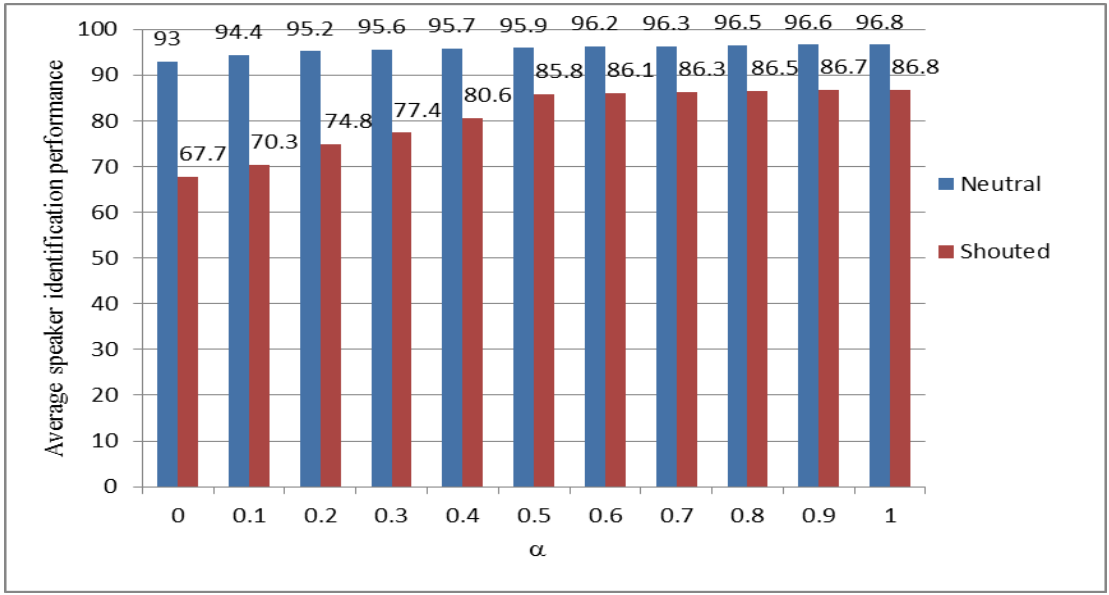

**Figure 2.** "Average speaker identification performance in each of neutral and shouted environments" using the Emirati-accented dataset based on CSPHMM3s for various values of $\alpha$

6) Experiment 6: In this experiment, "a statistical cross-validation technique" has been performed to estimate the standard deviation of speaker identification performance in each of neutral and shouted environments using the Emirati-accented database based on each of CSPHMM1s, CSPHMM2s, and CSPHMM3s. Cross-validation technique has been independently carried out for each classifier as follows: the entire database (5400 utterances per classifier) has been randomly subdivided into five subsets per classifier. Each subset is made up of 1080 utterances (360 utterances have been used for training and the rest have been used for testing). The standard deviation has been calculated using these five subsets per classifier. The standard deviation values per classifier are given in Fig. 3. Based on this figure, cross-validation technique reveals that the computed standard deviation values are low. Low standard deviation values indicate that the attained values of speaker identification performance are homogenous and not much difference among values. Therefore, it is apparent from this experiment that speaker identification performance in each of neutral and shouted environments using the Emirati-accented dataset based on each one of these classifiers and using the five subsets is very alike to that using the whole database without portioned it arbitrarily into five subsets per classifier (very minor variations).

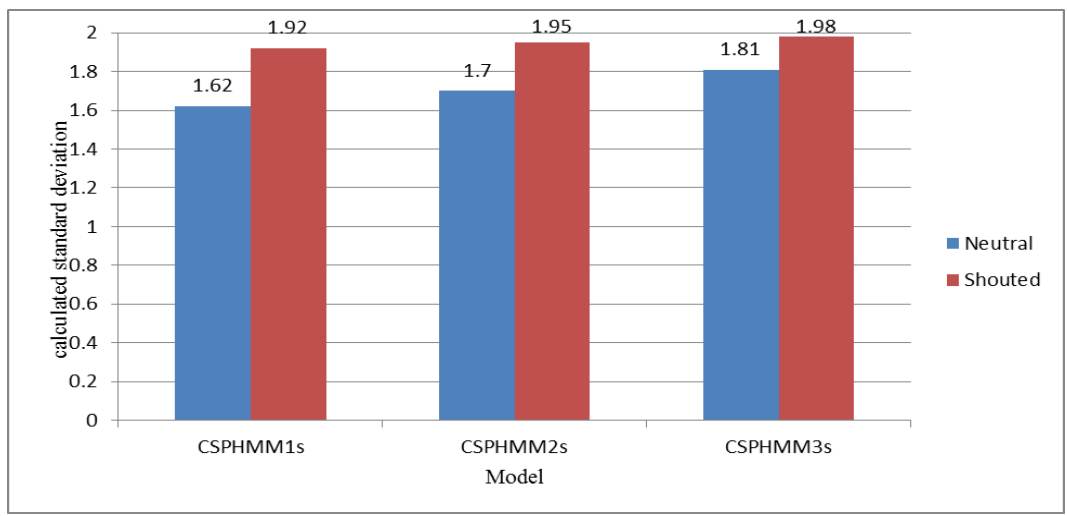

**Figure 3.** "Calculated standard deviation values using statistical cross-validation technique in each of neutral and shouted environments" of the Emirati-accented dataset based on each of "CSPHMM1s, CSPHMM2s, and CSPHMM3s"

7) Experiment 7: An "informal subjective assessment of CSPHMM3s" using the Emirati-accented dataset has been conducted with ten "nonprofessional listeners (human judges)". A total of 800 utterances (50 speakers × 8 sentences × 2 talking environments) have been utilized in this assessment. During the assessment, every listener was independently asked to recognize the unknown speaker in each of neutral and shouted environments (totally two distinct and separate environments) for every test utterance. The "average speaker identification performance in neutral and shouted environments" based on the subjective assessment is 93.9% and 56.8%, respectively. These averages are very similar to the reported averages in the present study based on CSPHMM3s (95.9% and 59.3% in neutral and shouted environments, respectively).

## 7. Concluding Remarks

In this work, Emirati-accented speech database (Arabic United Arab Emirates database) was captured in each of neutral and shouted environments in order to study, analyze, and improve text-independent Emirati-accented "speaker identification performance in each of neutral and shouted environments" based on each of "CSPHMM1s, CSPHMM2s, and CSPHMM3s" as classifiers. These classifiers are novel for such a database. In this study, seven extensive experiments have been carried out to thoroughly study and analyze this database for "speaker identification in each of neutral and shouted environments". Some conclusions can be drawn in this work. Firstly, as classifiers, "CSPHMM3s" outperform each of "GMMs, SVM, GA, VQ, CSPHMM1s, and CSPHMM2s" for "speaker identification in each of neutral and shouted environments". Secondly, in terms of gender, male Emirati speakers can be easily identified compared to female Emirati speakers. Thirdly, suprasegmental models yield higher speaker identification performance than their corresponding acoustic models. Finally, speaker identification performance based on training and testing speaker identification system using Emirati-accented database is greater than that based on training and testing the system using formal-accented Arabic database.

Our work has some limitations. First, the collected dataset is limited to a total of fifty speakers. Second, the speakers are unprofessional ones. Third, there are some logistic problems in capturing speech signals from senior people. Finally, MFCCs have been adopted in this work as the proper features that extract the phonetic content of Emirati-accented speech database.

For future work, our plan is to make our captured Emirati-accented speech database a comprehensive one by including more speakers. In addition, we plan to include speakers from early ages (5 to 12 years old) and to include speakers from the seven emirates of the UAE (Abu Dhabi, Dubai, Sharjah, Ajman, Umm al-Qaiwain, Ras al-Khaimah, and Fujairah). We also intend to determine the optimum extracted features that can be adopted as the appropriate features to extract the phonetic content of Emirati-accented speech signals. Finally, we plan to use deep neural networks [38] as classifiers to enhance Emirati-accented "speaker identification performance in a shouted environment".

## Acknowledgements
The authors of this work wish to thank University of Sharjah for funding their work through the competitive research project entitled "Capturing, Studying, and Analyzing Arabic Emirati-Accented Speech Database in Stressful and Emotional Talking Environments for Different

Applications", No. 1602040349-P. The authors wish also to thank engineers Merah Al Suwaidi, Deema Al Rais, and Hannah Saud for capturing the Emirati-accented speech database.

**"Conflict of Interest**: The authors declare that they have no competing interests".

**"Informed consent**: This study does not involve any animal participants".

[18] I. Shahin and Mohammed Nasser Ba-Hutair, "Talking condition recognition in stressful and emotional talking environments based on CSPHMM2s," International Journal of Speech Technology, Vol. 18, issue 1, March 2015, pp. 77-90, DOI: 10.1007/s10772-014-9251-7.

[19] I. Shahin, "Studying and enhancing talking condition recognition in stressful and emotional talking environments based on HMMs, CHMM2s and SPHMMs," Journal on Multimodal User Interfaces, Vol. 6, issue 1, June 2012, pp. 59-71, DOI: 10.1007/s12193-011-0082-4.

[20] T. S. Polzin and A. H. Waibel, "Detecting emotions in Speech," Cooperative Multimodal Communication, Second International Conference 1998, CMC 1998.

[21] I. Shahin, "Speaker identification in the shouted environment using suprasegmental hidden Markov models," Signal Processing Journal, Vol. 88, issue 11, November 2008, pp. 2700-2708.

[22] I. Shahin, "Employing second-order circular suprasegmental hidden Markov models to enhance speaker identification performance in shouted talking environments," EURASIP Journal on Audio, Speech, and Music Processing, Vol. 2010, Article ID 862138, 10 pages, 2010. doi:10.1155/2010/862138.

[23] C. Zheng and B. Z. Yuan, "Text-dependent speaker identification using circular hidden Markov models", IEEE International Conference on Acoustics, Speech and Signal Processing, S13.3, 1988, pp. 580-582.
35